\documentclass{article}

\usepackage{PRIMEarxiv}

\usepackage[utf8]{inputenc} 
\usepackage[T1]{fontenc}    
\usepackage{hyperref}       
\usepackage{url}            
\usepackage{booktabs}       
\usepackage{amsfonts}       
\usepackage{nicefrac}       
\usepackage{microtype}      
\usepackage{lipsum}
\usepackage{fancyhdr}       
\usepackage{graphicx}       
\graphicspath{{media/}}     
\usepackage{array}
\usepackage{booktabs}
\usepackage{subfigure}
\usepackage{parskip}

\title{Multi-modal expressive personality recognition in data non-ideal audiovisual based on multi-scale feature enhancement and modal augment
}

\author{
  Weixuan Kong, Jinpeng Yu, Zijun Li, Hanwei Liu, Jiqing Qu, Hui Xiao, Xuefeng Li \\
  Tongji University \\
  \texttt{lixuefeng@tongji.edu.cn} 
}

\begin{document}
\maketitle

\begin{abstract}
Automatic personality recognition is a research hotspot in the intersection of computer science and psychology, and in human-computer interaction, personalised has a wide range of applications services and other scenarios. In this paper, an end-to-end multimodal performance personality is established for both visual and auditory modal datarecognition network , and the through feature-level fusion , which effectively of the two modalities is carried out the cross-attention mechanismfuses the features of the two modal data; and a is proposed multiscale feature enhancement modalitiesmodule , which enhances for  visual and auditory boththe expression of the information of effective the features and suppresses the interference of the redundant information. In addition, during the training process, this paper proposes a modal enhancement training strategy to simulate non-ideal such as modal loss and noise interferencedata situations , which enhances the adaptability ofand  the model to non-ideal data scenarios improves the robustness of the model. Experimental results show that the method proposed in this paper is able to achieve an average Big Five personality accuracy of  , which outperforms existing 0.916 on the personality analysis dataset ChaLearn First Impressionother methods based on audiovisual and audio-visual both modalities. The ablation experiments also validate our proposed , respectivelythe contribution of module and modality enhancement strategy to the model performance. Finally, we simulate in the inference phase multi-scale feature enhancement six non-ideal data scenarios to verify the modal enhancement strategy's improvement in model robustness.
\end{abstract}


\section{Introduction}

Personality has long been the of psychologists' researchfocus , and it is a comprehensive expression of individual psychological traits that permeate human cognition, emotion and behaviour. From a psychological perspective, personality not only reflects a person's specific response patterns in a stable environment, but also embodies the complex interweaving of his or her intrinsic motivations, values and worldview. From another perspective, the influence of personality is ubiquitous in human society, cutting across a wide range of domains, including interpersonal relationships, career choices, educational strategies, and mental health. For example, people who are more extroverted are more inclined to engage in occupations that require frequent social interactions, while people who are more dutiful usually show higher levels of self-discipline and task completion. In addition to this, first impressions of an unfamiliar face can be generated in less than 100 milliseconds of contact based on personality traits\cite{willis2006first} , and first impressions play an important role in everyday life, such as interviews, elections, and dating. Various approaches and theories have been developed to classify, explain and measure personality. Vinciarelli et al. argue that personality is a psychological construct whose main purpose is to explain the diversity of human behaviour and which \cite{vinciarelli2014survey}can be predicted by stable and measurable characteristics of individual human beings. Costa et al. argue that traits are the key elements in the make-up of personality, and are the used to \cite{costa1998trait}basic units of personalitymeasurement measure and assess , and argue that human habitual patterns of and emotions are relatively stable over time. behaviour, perceptions As psychologists have progressed in their research on personality, several different models of personality have evolved, such as the Big Five Model\cite{mccrae1992introduction} , the Cattell 16 Personality Factors (16PF)\cite{karson1976guide} , the MyersBriggs Type Indicator (MBTI)\cite{furnham1996big} , the Minnesota Multi-Personality Inventory (MMPI)\cite{graham1978minnesota} , and the Eysenck Personality Questionnaire (EPQ)\cite{hamilton1977eysenck} , among others. Among them, the Big Five model is considered to be one of the most influential models in personality expression, which represents personality through five dimensions: openness, conscientiousness, extraversion, pleasantness , and neuroticism, and the quantitative personality dimensions make personality research have far-reaching theoretical and practical significance in the field of psychology. However, no matter which personality model is used, mainly through questionnairespersonality is , which due to  large number of questions analysed leads to personality time-consuming and laborious identification and analysis theand objective factors.

In recent years, with the rapid development of artificial intelligence and big data technology, massive multimedia data are emerging in the Internet, which provides unprecedented opportunities for automatic personality recognition based on multimodal data\cite{junior2019first,mushtaq2023vision,ilmini2024detection} . Personality recognition technology aims to automatically infer the potential personality traits of an individual by analysing his/her multimodal information such as speech, voice and facial expression. Compared with traditional personality questionnaire assessment methods, this technology is able to capture the behavioural and psychological characteristics of an individual with higher efficiency and accuracy, providing rich application prospects in a variety of fields such as psychology research, intelligent assistants, and social robots. Especially in intelligent interaction systems, it can identify the user's personality and emotional state in real time through multimodal information, and then provide personalised services and feedbacks to significantly enhance the user experience.

However, despite the large amount of current research devoted to improving the accuracy of multimodal personality recognition ,there are still several pressing challenges that need to be addressed. Firstly, effective mining and fusion of multimodal data is a core issue in improving recognition accuracy. Due to the differences between different modalities, it is still a complex and challenging task to extract effective features from different modalities through appropriate technical means, and to combine these modal features organically so as to give full play to their complementary characteristics.  Secondly, in practical applications, modal loss and noise effects are inevitable, which puts higher requirements on the stability and robustness of recognition results. How to maintain efficient and accurate recognition in the presence of partial modal loss or noise interference has become a key problem in the field of multimodal personality automatic recognition. Therefore, future research should focus on how to optimise the modal feature extraction and fusion algorithms to enhance the adaptive ability of the model in complex environments, so as to promote the widespread application of this technology in practical applications.

This paper is dedicated to the study of audio-visual multimodal automated personality recognition algorithms, which not only aims to improve the automated understanding of personality, but also hopes to in such as mental health assessment, personalised services and social behaviour predictionprovide value applications . real-world The main contributions of are as follows:this research 

(1)	An proposedend-to-end audio-visual multimodal algorithmic framework for automatic personality recognition, which we call MsMA-Net , is , which strengthens the relevance and discriminative power of multimodal information by through the fusing visual and auditory modal information fusion of , and feature-level the cross-attention mechanismby deeply exploring and capturing the key interaction features between the visual and auditory modalities, and the proposed method accuracy, robustness, and generalisation ability outperforms the existing based on the visual and auditory modalities in terms of accuracy, robustness and generalisation abilityof other algorithms .fusion of 

(2)	A is proposed for both visual and auditory modalitiesmulti-scale feature enhancement module (MSFEM) , which suppresses the interference of redundant features and mines and enhances high-value features that are crucial to the personality recognition task through a multi-scale feature extraction mechanism, targeting the information characteristics of different modalities. The module is based on a multi-scale feature extraction mechanism.

(3)	A  is proposed modality augment strategy ( MAS ) , which enables the model to cope with the common in multimodal datamodal loss and modal noise , improves the accuracy and robustness of the model in non-ideal scenarios, and provides an effective solution to the key problems in multimodal learning, and at the same time lays a solid foundation for the popularisation of the multimodal technology in practical applications. It also lays a solid foundation for the popularisation of multimodal technology in practical applications.

\section{Related Work}

In recent years, personality auto-recognition has gained the attention of many scholars, and the existing personality auto-recognition tasks are mainly focused on the ChaLearn First Impression dataset released in the ECCV 2016 competition\cite{ponce2016chalearn} on , and a series of works still continue to refresh the list. On the other hand, since automatic personality recognition has applications in several domains, developing with high accuracy models and robustness is a natural hotspot. Earlier studies were mainly based on traditional machine learning methods, e.g., Ilmini et al. used Physiognomy method to \cite{ilmini2016persons}extract facial features and compared Support Vector Machines and Artificial Neural Networks, and found that Artificial Neural Networks were more effective.Qin et al.\cite{qin2016modern} extracted global and local features of the face by LBF, SIFT, Gist, and introduced Fingerprint Images together to predict personality. Various machine learning models such as Decision Tree, Random Forest, Plain Bayes, and Linear Regression were used, and it was found that facial features can effectively predict partial personality with higher accuracy for females than males. Traditional methods are certainly effective, however, the shortcomings of manually designing features and the low accuracy rate make the field develop slowly, with the development of deep learning technology, automatic personality recognition based on automatic extraction of features from deep networks develops rapidly and gradually replaces traditional machine learning methods.

Deep learning based automatic personality recognition studies are mainly classified into unimodal and multimodal, Hayat et al.\cite{hayat2019use} used a pre-trained audio feature extraction convolutional network and it on fine-tuned  to obtain an audio-based automatic personality recognition model.the First Impression datasetZhu et al.\cite{zhu2018automatic} improved the LSTM algorithm to augment the data by frame-skipping sampling, and the model can automatically extract audio from MFCC features Mandarin speech, from Mandarin speech the model can automatically extract audio MFCC features to identify personality traits.Ventura et al.\cite{ventura2017interpreting} used a modified convolutional network DAN to extract facial features and visualised the convolutional layers using CAM, and found that the eyes, nose, mouth of a person and eyebrows are closely related to their personality traits.. Gürpınar et al.\cite{gurpinar2016combining} extracted facial expression and surrounding environment features a VGG-19 network, and then merged the two features togetherfeatures, by using pretrained after which the two features were merged and fed into a kernel-limit learning machine to on predict Big Five personality . the First Impression datasetAlthough unimodal-based personality recognition has achieved some results, with the technology updating, multimodal tasks have gradually gained attention. Often, multimodal data can complement each other's information to obtain a more accurate and robust model, so multimodal automatic personality recognitiona large number of research results emerge in . have begun to Zhang et al. used \cite{zhang2016deep}pre-trained VGGFace to design DAN and DAN+ architectures for visual modal feature extraction, and a logarithmic filter bank to extract audio features, which were then fused together to participate in the personality prediction, and their work won the ECCV competition.Güçlütürk et al.\cite{guccluturk2016deep} developed an audio-visual residual network using a 17-layer residual network for feature extraction in both modalities, before fusing the features of both and predicting the Big Five personality.Kampman et al.\cite{kampman2018investigating} performed personality prediction through the three modalities of visual, auditory, and textual modalities, specifically, using a pre-trained VGGFace to extract visual features, a convolutional network to extract audio features, and Word2vec to extract text features, and the results show that visual modalities are more important for personality prediction than audio and text modalities. Li et al. also used the three modalities and proposed a three-branch classification-regression network\cite{li2020cr}, CR-Net, where the first branch is used for visual features, the second for facial features, and the third branch for audio and their features, the first branch for visual features, the second for facial features, and the third for audio and its transcribed text, with each branch ending with a regressor for predicting Big Five personality scores. Song et al. first \cite{song2021self}trained a generic U-net model to learn general facial dynamics from unlabelled videos of faces, and then further self-supervised learning by using only person-specific videos to obtain individualised facial dynamics features to be used for predicting personality traits. Principi et al.\cite{principi2019effect} point out that existing automatic personality perception methods do not directly predict the true personality of the target, but rather the extrinsic personality assigned by an external observer, and therefore need to cope with the human bias inherent in the training data. Based on this, they investigate different sources of bias affecting personality perception, such as emotion in facial expressions, attractiveness, age, gender, and ethnicity, and explore their impact on the predictive power of extrinsic personality estimation. To this end, a multimodal deep neural network was proposed to combine raw audio and visual information with predictions from attribute-specific models for regression of Big Five personality traits.Bohy et al.\cite{10581940} proposed Social-MAE, a based on an extended version of the Contrastive Audio-Video Masked Self-Encoder (CAV-MAE)pre-trained audio-video model , for perceiving human social behaviour. It mainly improves CAV-MAE to handle more frames as inputs and performs on a large-scale social interaction dataset (VoxCeleb2)self-supervised pre-training , followed by predicting personality on a downstream task.Duan et al.\cite{10674375} proposed a multimodal extrinsic personality analysis method by combining a modified ConvNeXt network with gated recurrent units (GRUs) to analyse visual modalities in short videos, and using a random forest model to predict audio features. In order to effectively fuse the information of audio and visual modalities, an ECA attention mechanism and a new network structure unit are introducedECblock .Shen et al.\cite{shen2024pacmr} proposed a progressive adaptive cross-modal reinforcement method (PACMR) for multimodal extrinsic personality characterisation, which aims to solve the problem of asynchrony between modalities.PACMR achieves multilevel information exchange between modalities through progressive reinforcement strategies, thus simultaneously enhancing the source modalities. information exchange, thus enhancing the features of both source and target modalities simultaneously.Sun et al.\cite{10386376} proposed a novel Multimodal Co-Attention Transformer network (MCoAttention Transformer) for video-based emotion prediction. The method overcomes existing methods in modal by simultaneously modelling audio, visual and textual representations and their interrelationshipsthe limitations of terms of high computational cost , enabling efficient and accurate personality prediction.interaction and Zhang et al.\cite{8897617} proposed PersEmoN for joint learning of affective and extrinsic personality features from facial images. The network is trained separately for affective and extrinsic personality datasets by sharing the underlying feature extraction module and optimising it under a multi-task learning framework. In addition, an adversarial-like loss function is used to promote consistency of representation . across datasetsIn addition to these research papers, there are also scholars who have made great contributions to the field of automatic personality recognition, Liao et al.\cite{liao2024open} proposed the first reproducible audio-video extrinsic personality computational benchmark for fairly and consistently evaluating eight existing computational models of personality (including audio, visual, and the performances of models) and seven standard deep learning models in self-report and extrinsic personality recognition tasks.audio-video 

In this paper, we will focus on two modalities, video and audio, and propose our multimodal personality recognition network MsMA-Net, and design a multiscale feature enhancement module that can be used for both modalitiesMSFEM , while the modal enhancement strategy is introduced during the training process MAS to help the model to cope with the non-ideal data scenarios. Compared with other works of the same type, our network  lower structural complexityhas, better performance, and higher robustness.

\section{Methods}
In this section, we will firstly introduce the overall framework of the proposed end-to-end multimodal representation of the automatic personality recognition algorithm, secondly introduce the proposed multiscale feature enhancement module for , and finally introduce the both visual and auditory modal datamodal enhancement strategy  .in the training phase

\subsection{Overall framework}
Our proposed algorithmic architecture employs two branches corresponding to the two modalities of video and audio, as shown in Fig. 1. In the training phase, the input data is first processed by a modal enhancement strategy, which is able to simulate possible non-ideal situations, such as modal loss, noise interference, etc., so as to improve the robustness of the model to imperfect data in the inference phase. Through this pre-processing step, the model can effectively cope with various complex situations in real applications to ensure its stability and reliability. Afterwards, the corresponding feature extraction will be performed for the two modalities respectively, and by the proposed feature enhancement will be performed generic feature extraction module. audio-visual Finally, the based on the cross-attention mechanism feature-level fusion module deeply fuses the features of visual and auditory modalities, which enables the model to capture important interaction information between different modalities, thus realising the complementarity of modal information, which helps to enhance the discriminability of the features, and improves the model's discriminative ability and recognition accuracy.

\begin{figure}[htbp]
  \centering
  \includegraphics[width=1\textwidth]{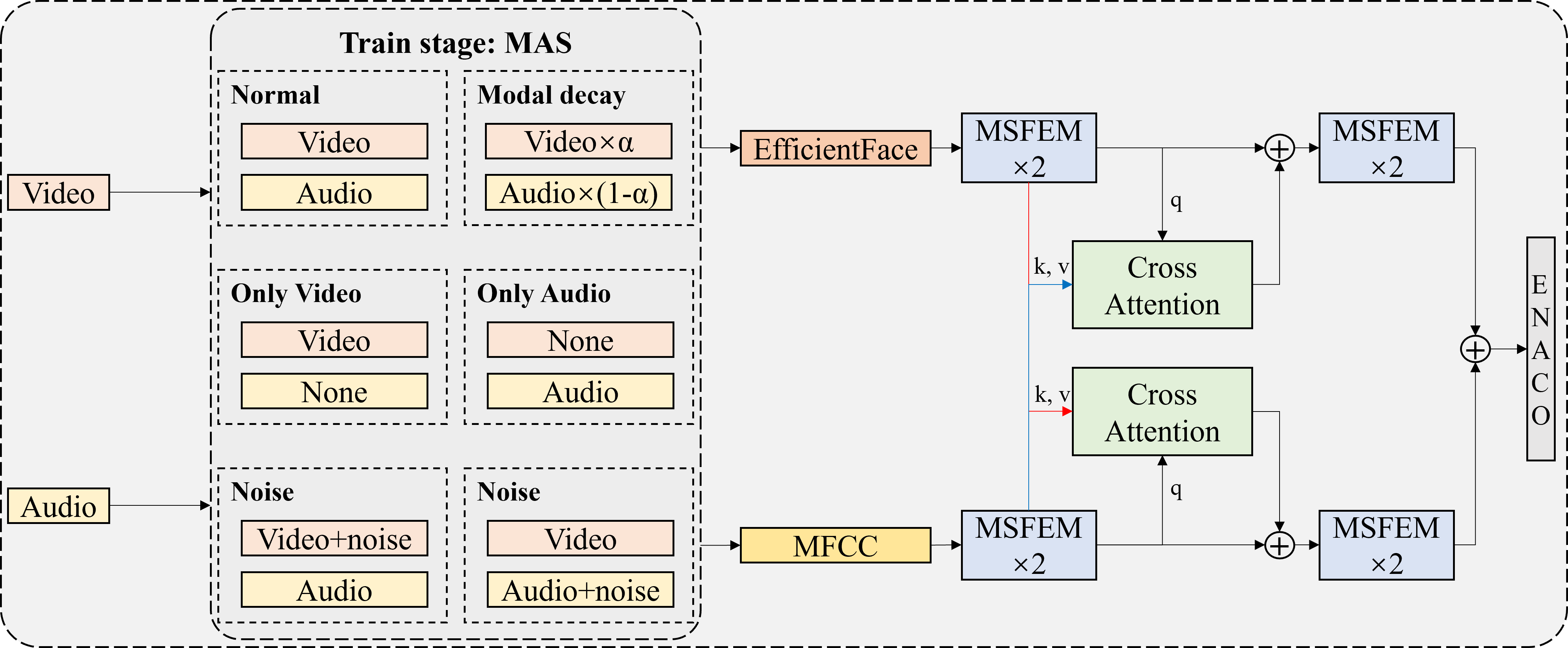} 
  \caption{MsMA-Net overall architecture}
  \label{fig:demo}
\end{figure}

For the visual branch, we extracted features from multiple frames in the video to construct a feature representation of the entire video. In order to improve the accuracy of the visual branch feature representation, we used the pre-trained EfficientFace model as a feature extractor. Although the EfficientFace model was initially used for the expression recognition task, it has demonstrated a strong capability in facial feature extraction, which is crucial for multimodal personality recognition. Therefore, with the help of this model, we are able to more accurately characterise the visual features in the video and improve the overall model recognition. For the auditory branch, we represent the audio features by extracting the MFCC (Mel-Frequency Cepstral Coefficients) features of the whole audio, which is a widely used feature in speech recognition that can effectively reflect the spectral characteristics of audio signals, and is especially suitable for processing auditory information related to emotion and personality.

Compared with unimodal, multimodal information has a natural complementarity in feature expression, which can effectively enhance the robustness and accuracy of the model. Therefore, in order to take full advantage of the complementary nature of multimodality, we choose to perform through the cross-attention mechanismfeature-level fusion .  traditional Unlikefusion or simple splicing, modal-level feature-level fusion can deeply integrate multimodal features to effectively capture the interaction information between different modalities. Moreover, the cross-attention mechanism can automatically adjust and weight the contribution of each modal feature by learning the relationship between different modalities, thus reducing the interference of redundant and useless information and focusing on those features with higher discriminative power. Specifically, we use two cross-attention mechanism modules for fusing the feature information of the two modalities, the first module will calculate the of visual and auditory features attention scores and obtain the weighted feature $F_{fusion1}$, and the second module will calculate the of auditory and visual features attention scores and obtain the weighted feature $F_{fusion2}$. Assuming that $F_{v}$ and $F_{a}$ represent visual and auditory features respectively, and $W_{q}$, $W_{k}$, and $W_{v}$ represent the calculating respectively, weight matrices for the Query, Key, and Value matrices, then the two cross-attention mechanisms can be expressed as Equations (1) and (2):

\begin{center}
  $F_{fusion1}=softmax \left ( \frac{F_{v}W_{q}\left ( F_{a}W_{k}  \right )^{T} }{\sqrt{d_{k}} }  \right )F_{a}W_{v} $

$F_{fusion2}=softmax \left ( \frac{F_{a}W_{q}\left ( F_{v}W_{k}  \right )^{T} }{\sqrt{d_{k}} }  \right )F_{v}W_{v} $
\end{center}

Taking $F_{fusion1}$ as an example, the query matrix comes from visual features, while the key matrix and the value matrix come from audio features. After the attention calculation, cross the visual and audio features are effectively fused, and the fused features not only contain the common features of the two modalities, but also realise the complementary information of the two modalities, which makes the features more discriminative.

\subsection{Audio-visual generic multi-scale feature enhancement module}

In order to improve the accuracy and multi-scale expression ability of features, we designed a multi-scale information fusion feature enhancement module, as shown in Fig. 2, which is able to adapt to both audio-visual and audio-visual modal data simultaneously, and effectively enhance the expression ability of features. The core idea of this module is to capture information at different scales through multi-scale convolutional operations, so as to enrich the hierarchy and details of the features, and thus improve the model's ability to understand complex input data. Specifically, the input features are first passed through a 1×1 convolutional layer, which is used to integrate the information and reduce the feature dimensionality. Next, we perform three different scales of convolution operations on the features: 1×1 convolution, 3×3 convolution, and 5×5 convolution, in order to capture the changes of the features in different spatial scales. Through this multi-scale convolution approach, the model can obtain multi-level feature representations ranging from local details to global information, which enhances the comprehension ability of the features at different levels and granularities. Afterwards, the features after 1×1 convolution will be spliced with the sum of the features after 1×1 and 3×3 convolution as well as the sum of the features after 3×3 and 5×5 convolution, respectively, and again by a 1×1 integrated convolution layer to ensure that the information at different scales can be effectively fused to avoid excessive redundancy and loss of information. Finally, in order to further enhance the feature recognition ability, we introduce a channel attention mechanism, which enables the model to automatically learn and enhance the attention to important feature channels, while suppressing the influence of redundant and useless information, and focusing on those features that are more discriminative for the personality recognition task, thus effectively improving the accuracy of feature representation, and consequently enhancing the performance of the overall model.
\clearpage

\begin{figure}[htbp]
    \centering
    \includegraphics[width=0.5\textwidth]{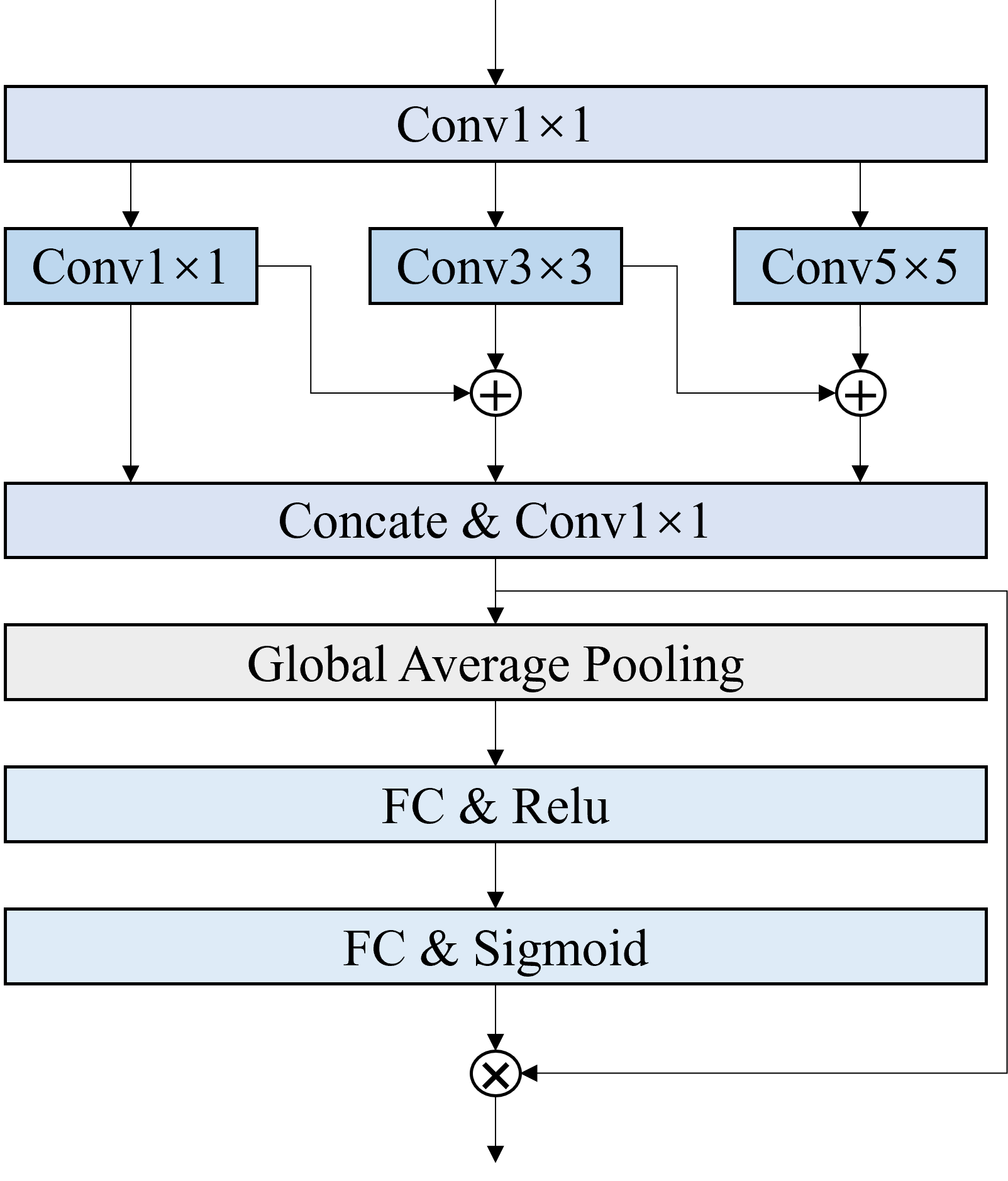}
    \caption{Schematic diagram of the multi-scale feature enhancement module}
      \label{fig:demo}
\end{figure}

\subsection{Modal Enhancement Strategy}
Real-world data acquisition environments are often characterised by complexity and uncertainty, and differ significantly from data acquisition under tightly controlled conditions in the laboratory. For example, due to malfunctioning acquisition equipment, environmental condition limitations, or temporary modal absence, the acquired data may contain only a single modality, or there may be significant noise and distortion in the datanon-ideal conditions such as . The presence of such modal loss or noise can significantly affect the performance and robustness of multimodal models, especially in real application scenarios. To address this problem, this study proposes a modal enhancement strategy during training to by improve the robustness and adaptability of the model non-ideal situations. Specifically, the strategy  the visual and auditory modalities of each set of video data separately during trainingsimulating processes, generating five additional derived data forms to construct more diverse training samples, and Table 1 describes each data form and its description in detail.

\begin{center}
Table 1 5 derived data under modal enhancement strategy

\begin{tabular}{lll}

  \toprule
  data format & instructions    \\
  \midrule
  Visual data normal, auditory data empty   & Simulation of complete loss of auditory modality  \\
  \midrule
  Hearing data is normal, visual data is \\empty  & Simulation of complete loss of visual modality   \\
  \midrule
  Visual data normal, auditory data plus \\noise   & Simulation of auditory modalities disturbed by noise  \\
\midrule
  Auditory data normal, visual data plus \\noise&Simulation of visual modalities disturbed by noise  \\ 
  \midrule
    Visual and auditory data are simultaneously \\attenuated by a random factor&Enhancing data diversity while modelling modal quality degradation \\

  \bottomrule
\end{tabular}
\end{center}

As shown in Table 1, each set of raw data will be expanded to contain six data forms of training data, including the raw data in ideal situations and the five non-ideal situationssimulated data in . These data share the same labels, and the of multimodal features during the training process joint learning helps the model dealing with non-ideal situationslearn to adaptively select effective modal features . This enhancement strategy allows the model to learn the feature representations under ideal conditions, but also makes full use of in non-ideal scenarioswhen the complementary nature of multimodal information , thus effectively improving the modelto modal loss or noise interference's adaptability .situations such as

\section{Experiment}
\subsection{dataset}
The dataset used in this study ChaLearn First Impression comes from the ECCV2016 competition , which contains a total of 10,000 video data, each video is about 15s in length, and the videos all come from YouTube, and the content is a person talking freely to the camera. All video data are divided into training set, validation set, and test set according to the ratio of 3:1:1, i.e., 6000 training data, 2000 validation data, and 2000 test data. Each piece of data was labelled with the Big Five personality trait labels of Extraversion, Conviviality, Responsibility , Neuroticism, and Openness, and each label as a in the interval [0, 1]was represented value continuous , and a partial sample of the data is illustrated in Figure 3. In order to with the competition have a uniform and fair comparison , all the models in this paper are trained using the training set only, and the best performing model is selected based on the model's results on the validation set, and its results on the test set are recorded.as well as the latest research

\begin{figure}[htbp]
    \centering
    \includegraphics[width=0.75\textwidth]{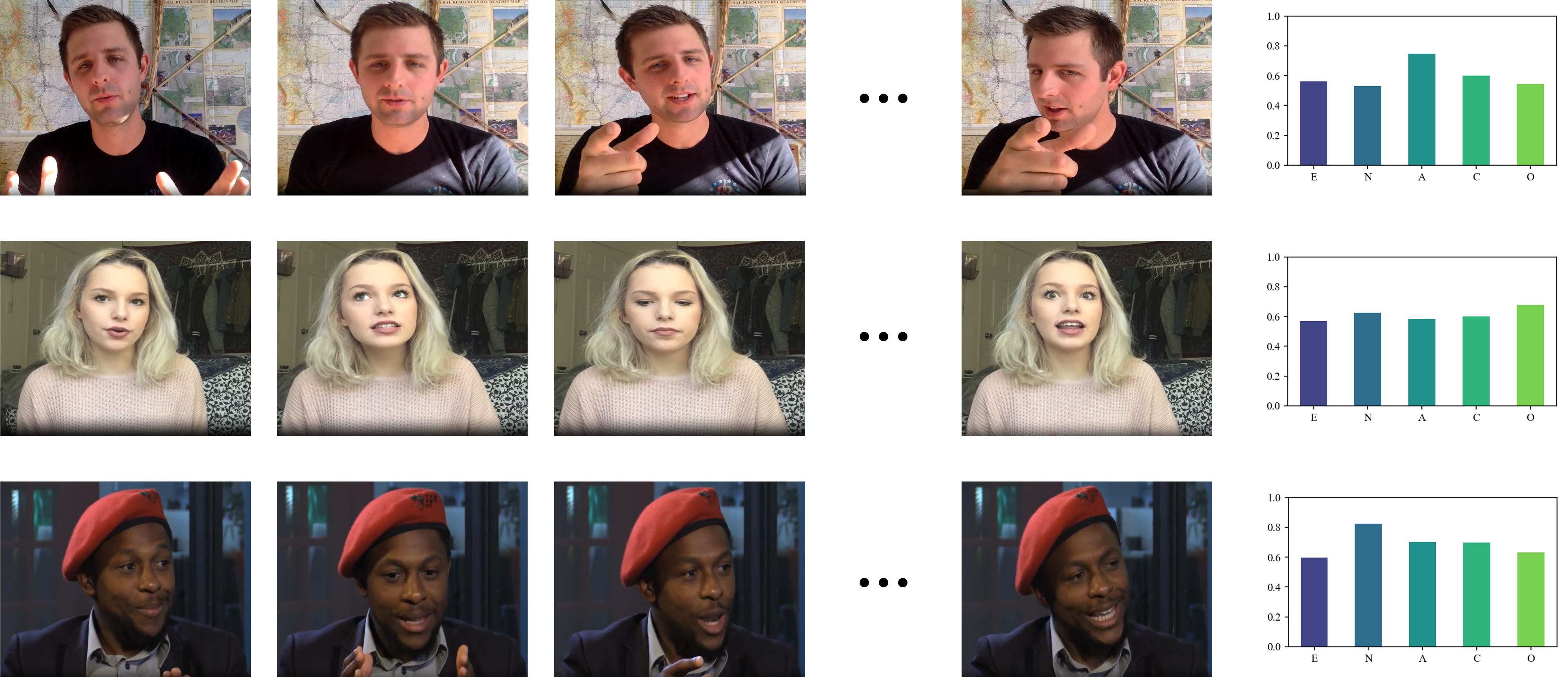}
    \caption{Some sample data}
\end{figure}

The evaluation metrics, which we use for calculations that are common on this task and widely used in competitions and subsequent studies, can be expressed as follows:

\begin{center}
$Accuracy_j=\frac{1}{N} \sum_{i=1}^{N} \left ( 1-\left | label_{ij}-predict_{ij} \right |  \right ) $       (3)
\end{center}

Here $label_{ij}$ represents the jth of thepersonality trait  ith video label, $predict_{ij}$ represents the jth videopersonality trait predicted value of , N represents the total number of videos, and j represents the number of personality traits, which takes the value of 1-5, representing the five personality traits of ENACO, respectively.the ith 

\begin{center}
$Average Accuracy=\frac{1}{p} \sum_{j=1}^{p} \left ( Accurracy_{j} \right ) $(4)
\end{center}

The five personality trait scores of ENACO can be obtained separately , and them from Eq. (3)averaging gives the average score, which also represents the overall performance of the model, as shown in , where p = 5.Eq. (4)

\subsection{ Experimental setup}

The hardware environment used in this study is centos7 operating system, two NVIDIA A40 graphics cards, software environment is python3.9, torch1.9.0, hyperparameters are set to batchsize of 8, random seed of 1, to ensure the reproducibility of all the experiments, the initial learning rate is 0.04, there are a total of training70 rounds of , and in the 30th, 45th, 55th, 60th round when the learning rate drops , respectivelyto previous oneone-tenth of the . All video data were pre-processed into two modal formats, visual and auditory, for auditory modality, we extracted 10 MFCC features; for visual modality, 75 frames of uniformly distributed images were taken out from the original 15s video and facial regions were extracted by a face detection algorithm, and finally all images were uniformly scaled to 224*224 size. When the model is trained, data from the same video data in both modal formats are input to the model simultaneously as a group, where the visual modality will be first by on the extracted model to extract featuresthe EfficientFace pre-trained AffectNet dataset .

\subsection{ Experimental results}

In this section, we will firstly present a our proposed algorithmcomparison of the results of algorithms  MsMA-Net with those of other existing modal based on both audio-visual; secondly, we will present the robustness of our model in the face of non-ideal data scenarios under the auspices of a modal enhancement strategy; and finally, we will present, through ablation experiments, the contributions of our proposed multiscale feature enhancement module and modal enhancement strategy, respectively.

As Table shown in , 2we compare at the time of the competition the results of the three and some recent studies, and we reproduce a network structure for emotion recognition and modify it so that it can be applied to the task of personality recognition, and top rankings we call it Baseline is designed by borrowing the idea of this structure.because our MsMA-Net The bolded fonts in represent the optimal results, and we can see that we achieve the optimal results in all the other four personality traits and the average index. We can see that except for Table 2 agreeableness, we achieve the optimal results in the other four personality traits and the average index, which can also show our the effectiveness and accuracy of algorithm.proposed  In addition, we can find that the Baseline results are not optimal, but they are still competitive, which lays the foundation for subsequent optimisation.
\begin{center}
Table 2 Comparison of results with other studies

\begin{tabular}{lllllll}

  \toprule
 Method &E &N&A&C&O& Aver  \\
  \midrule
   Zhang\cite{zhang2016deep} (2016)&0.913&0.910&0.913&0.917&0.912& 0.913  \\
  \midrule
 Güçlütürk\cite{guccluturk2016deep} (2016)&0.911&0.909&0.910&0.914&0.911& 0.911  \\
  \midrule
  Duan\cite{10075178} (2022) &0.888 &0.883&0.890&0.899& 0.894& 0.891\\
\midrule
Sun\cite{10386376} (2023)& 0.902	& 0.900	& 0.906	& 0.904	& 0.903& 	0.903\\ 
  \midrule
 Bohy\cite{10581940} (2024)&0.895	&0.905	&0.907&	0.902&	0.908	&0.903\\
  \midrule
  Duan\cite{10674375} (2024)& 0.906	&0.912&	0.907	&0.906	&0.908&	0.908\\
    \midrule
  Shen\cite{shen2024pacmr} (2024)& 0.911&	0.909&	0.914&	0.914&	0.912&	0.912\\
    \midrule
 Baseline\cite{9956592} (2022) & 0.916&	0.911&	0.908&	0.908&	0.912&	0.911\\
    \midrule
  MsMA-Net (ours)& 0.920	&0.916&0.913	&0.918&	0.915&	0.916\\
  \bottomrule
\end{tabular}
\end{center}

Figure 4 demonstrates the robustness that the modality enhancement strategy brings to the model in non-ideal data scenarios, and we can see that in the six non-ideal scenarios we listed, the performance of the model trained with the modality enhancement strategy is significantly better than that of the model without the modality enhancement strategy. In addition, we can find that the visual modality is more important in the automatic personality recognition task according to each set of comparisons, for example, in both (a)(b)sets of comparisons , the performance after missing the visual modality is lower than that after missing the auditory modality; in both (c)(d)sets of comparisons , the addition of the visual random noise modality by is the most affected , especially in conforming to the standard normal distributionset of experiments , where the model performance is only 0.42848, however, with the assistance of the modal enhancement strategy, the performance is significantly improved and is able to reach 0.87090; in (c)in the two in (e)(f)sets of comparisons , the addition of random noise has a large impact on both visual and auditory modalities, and with the assistance of the modal enhancement strategy,  case of normal vision with to the input features a similar distribution the improvement is smaller inthan in the case of normal hearing, which suggests that vision the by itself can bring a greater contribution, while auditory itself contributes relatively less. In addition to this we counted the average performance of the model trained with the modal enhancement strategy versus the original model in six non-ideal data scenarios, as Table shown in , and it can be seen that the model with the modal enhancement strategy improves significantly on five personality traits, and the overall 3average performance is improved by 12\%.

\clearpage
\begin{center}

\begin{figure}[htbp]
    \centering
    \subfigure[Only Video]{
        \includegraphics[width=0.25\textwidth]{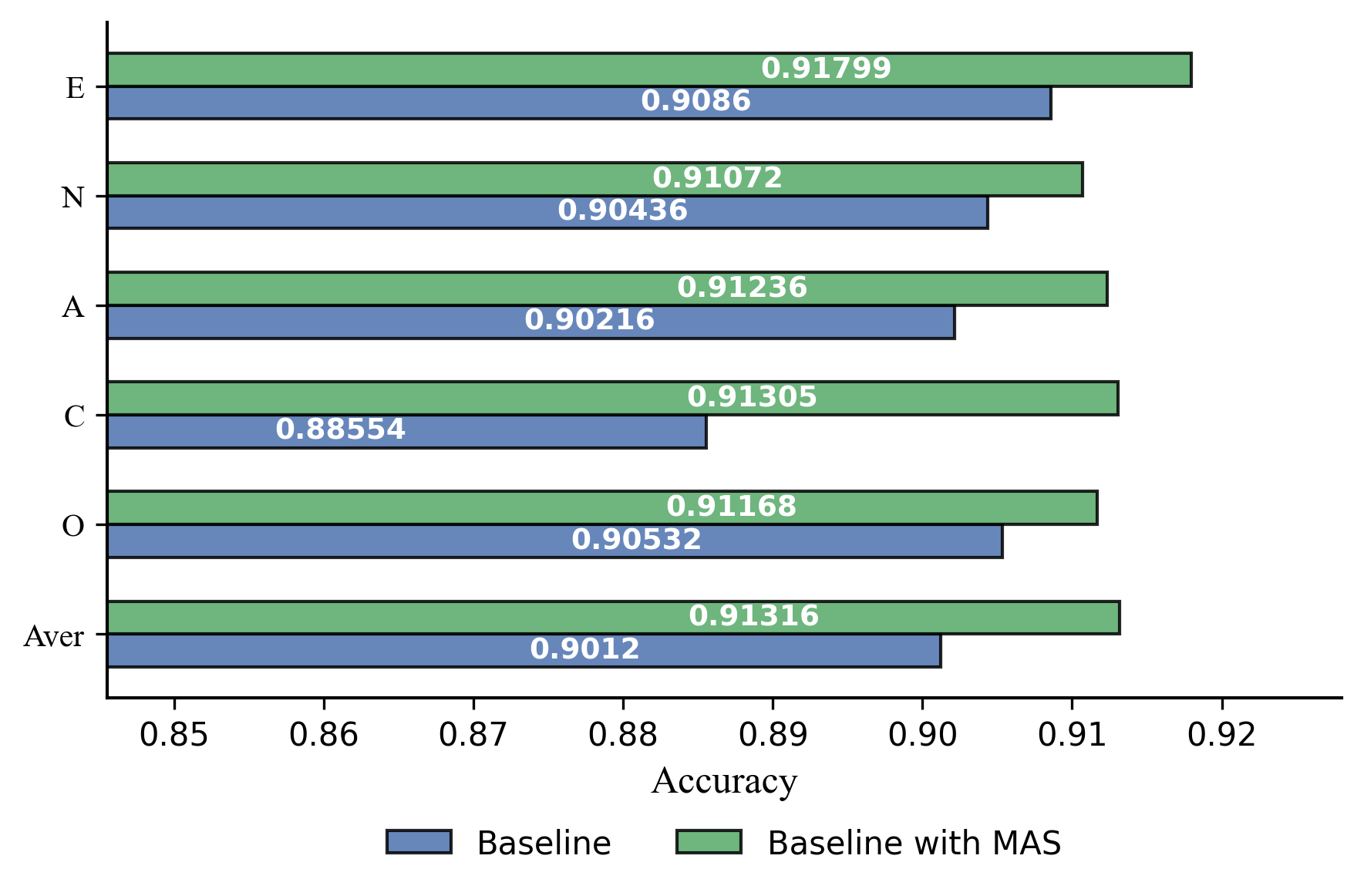}
    }
    \subfigure[Only Audio]{
	\includegraphics[width=0.25\textwidth]{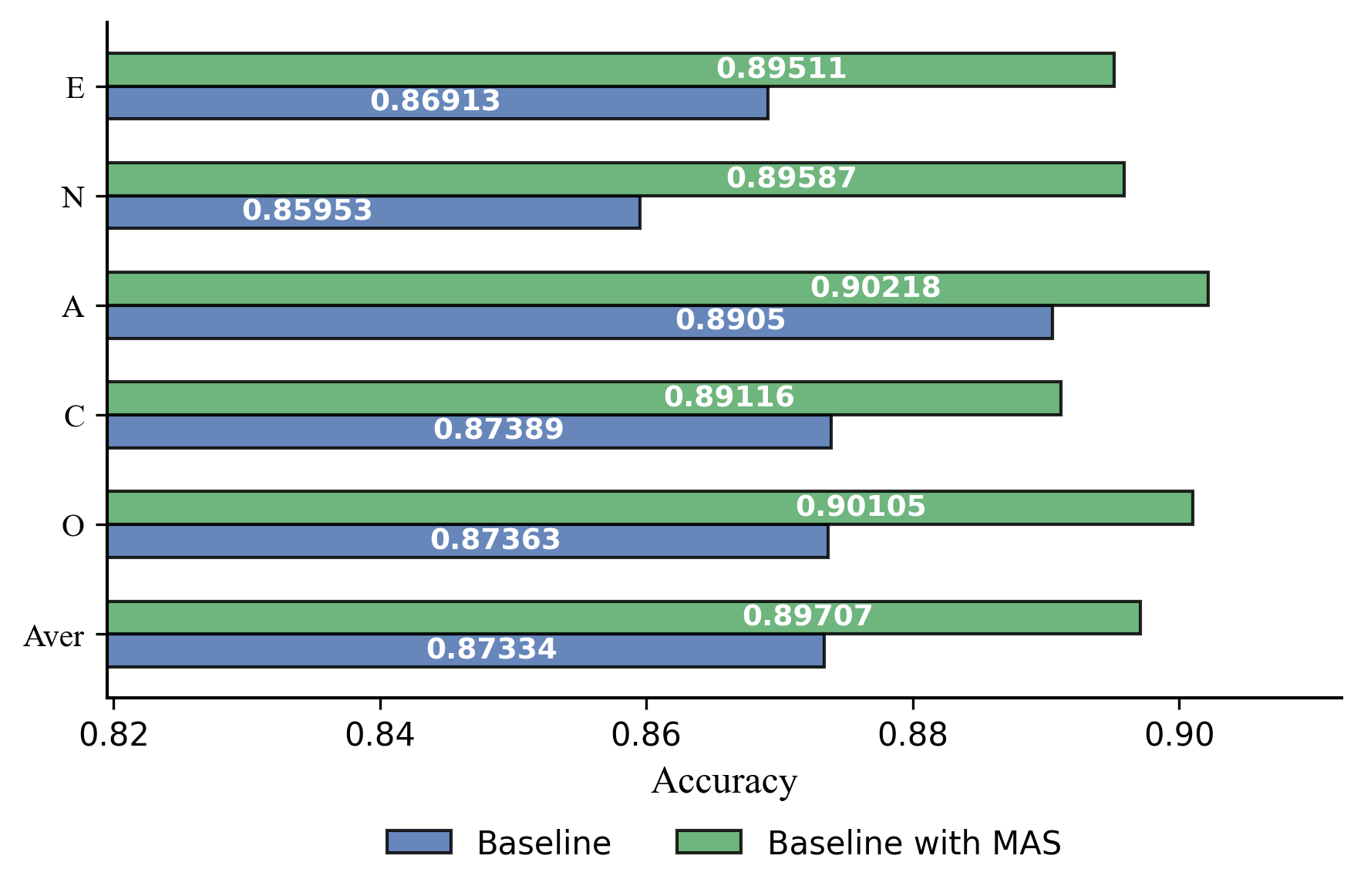}
    }
    
    \quad    
    
    \subfigure[Video w Full Noise]{
    	\includegraphics[width=0.25\textwidth]{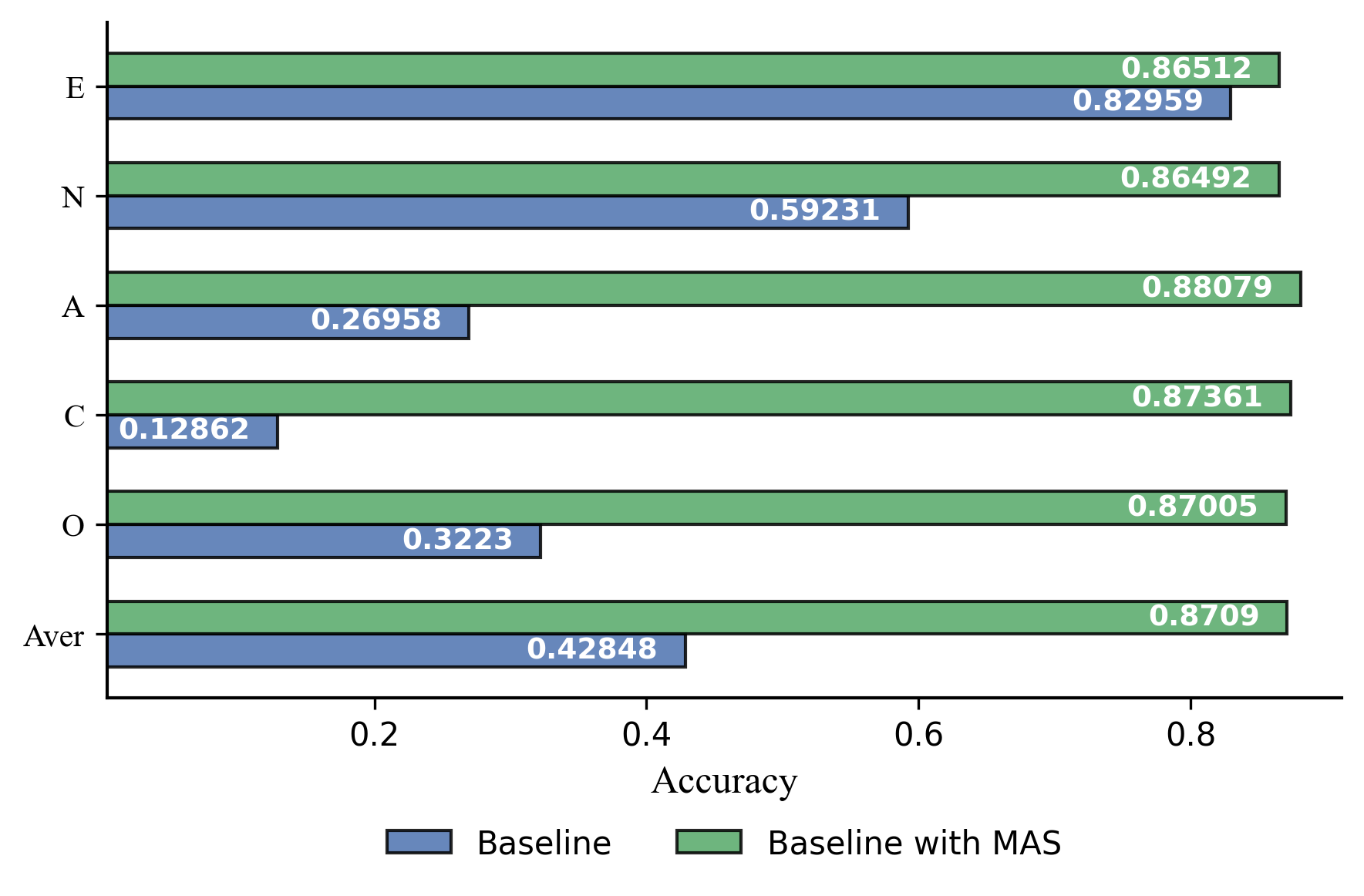}
    }
    \subfigure[Audio w Full Noise]{
	\includegraphics[width=0.25\textwidth]{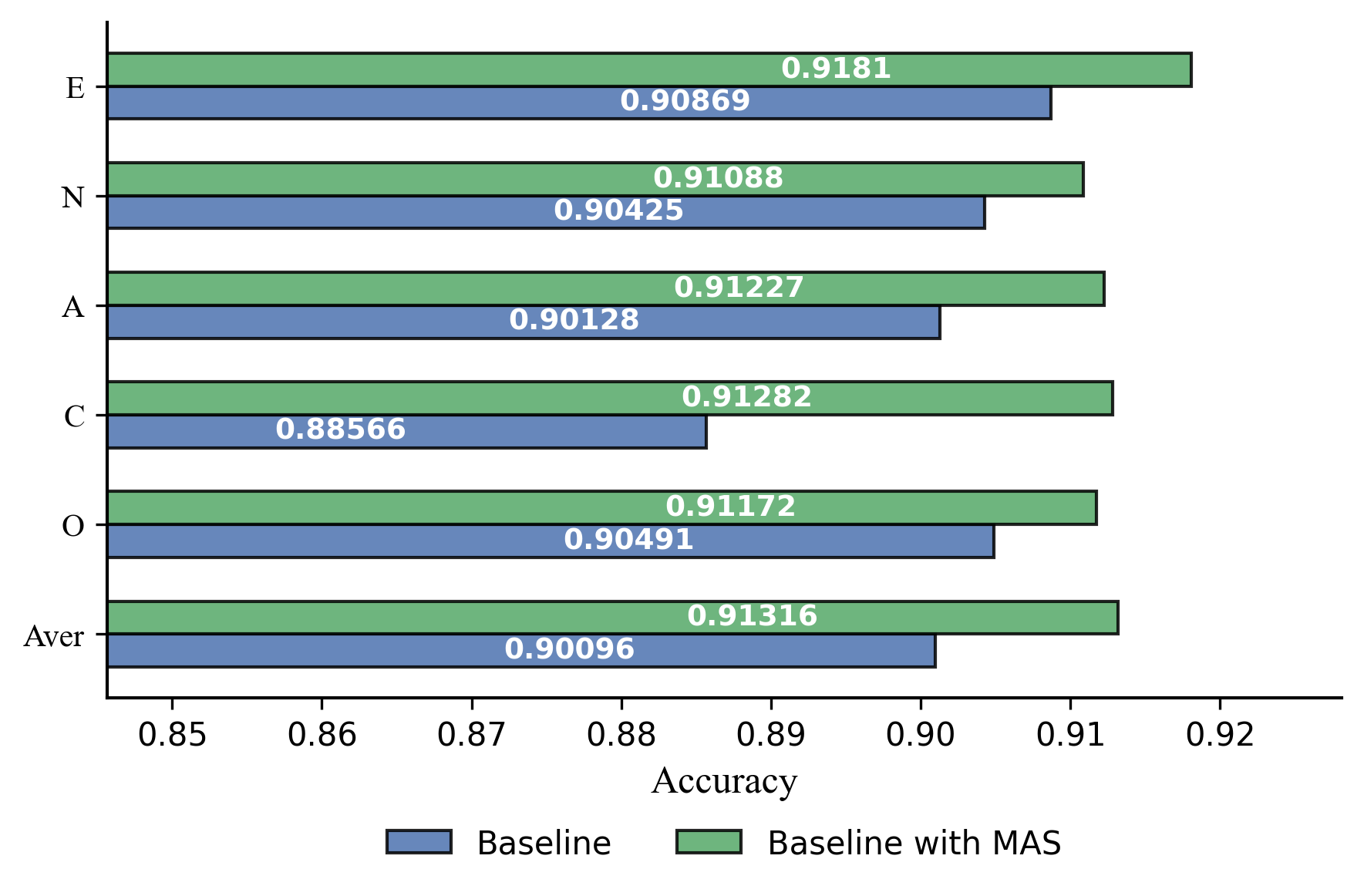}
    }
        
     \quad    
     
    \subfigure[Video w Noise]{
    	\includegraphics[width=0.25\textwidth]{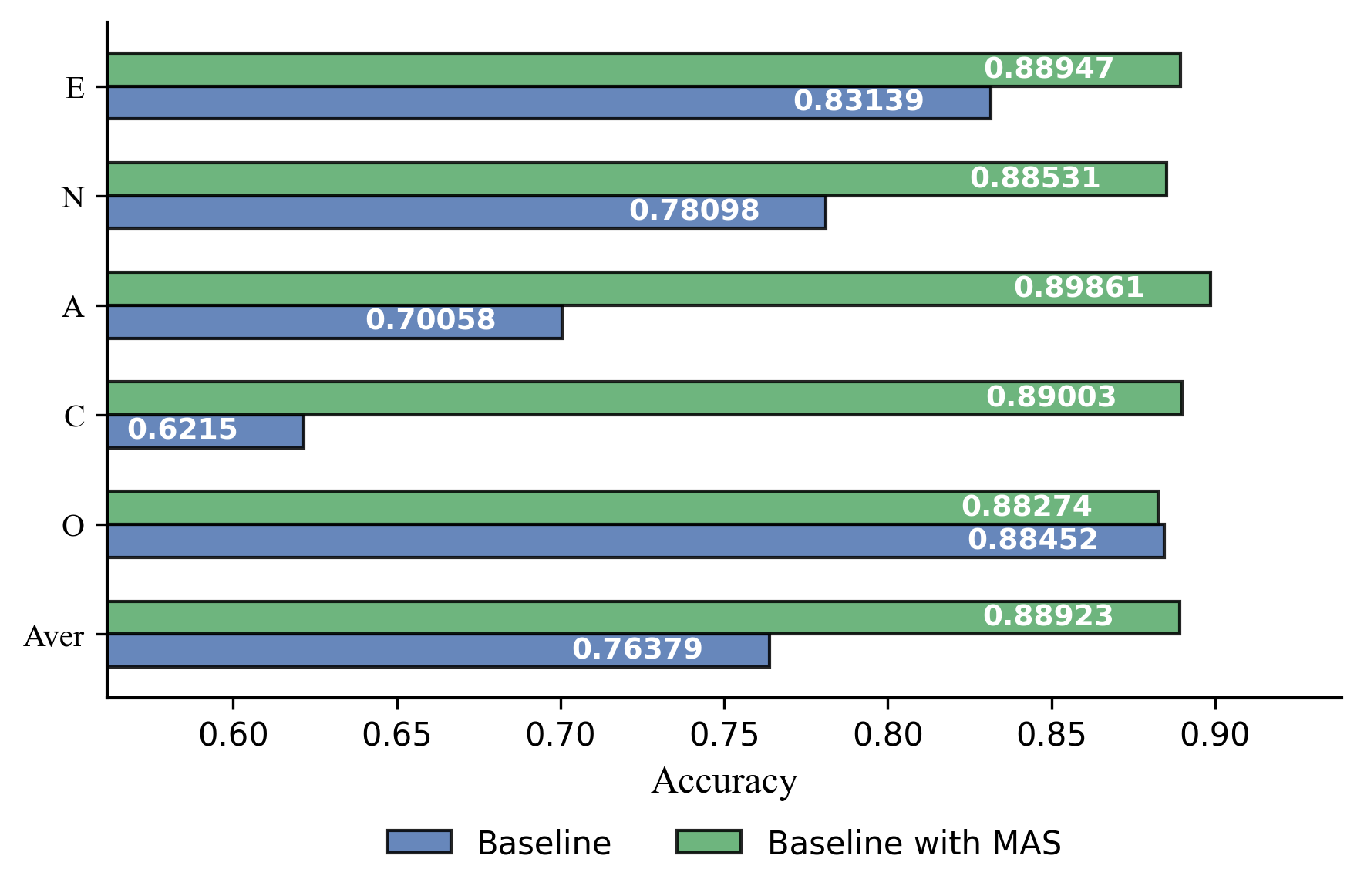}
    }
    \subfigure[Audio w Noise]{
	\includegraphics[width=0.25\textwidth]{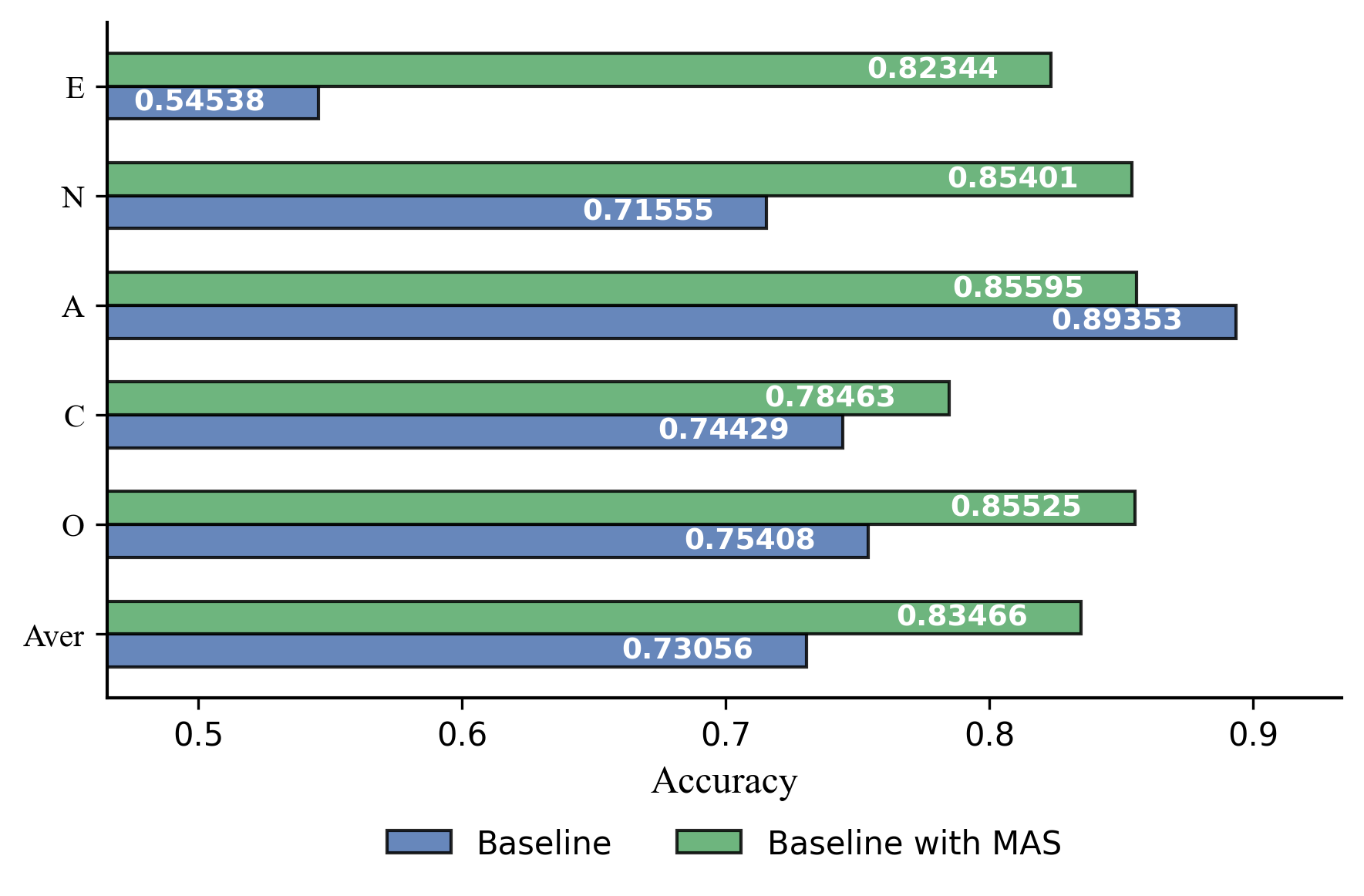}
    }
    \caption{ MAS results for six non-ideal data scenarios. (a) Visual modality only (b) Auditory modality only (c) Visual modality with random noise conforming to the standard normal distribution (d) Auditory modality with random noise conforming to the standard normal distribution (e) Visual modality with the addition of random noise similar to its distribution (f) Auditory modality with the addition of random noise similar to its distribution}

\end{figure}

\end{center}

\begin{center}
Table 3 Comparison of average performance of modal enhancement strategies in six non-
ideal data scenarios

\begin{tabular}{lllllll}

  \toprule
 Method &E &N&A&C&O& Aver  \\
  \midrule
  Baseline&0.81546&	0.79283&	0.75961&	0.68992	&0.77413	&0.76639  \\
  \midrule
Baseline w MAS&0.88487	&0.88695&	0.89369	&0.87755	&0.88875	&0.88636  \\

  \bottomrule
\end{tabular}
\end{center}

Table 4 demonstrates the results of the ablation experiments, our base framework with the addition of the multi-scale feature enhancement module as well as the modal enhancement strategy improves the average index by 0.2\% and 0.5\% respectively, and also improves on all five personality dimensions, especially on Conviviality and Responsibility, which verifies the validity of the innovative points we proposed. In addition the simultaneous combination of the multi-scale feature enhancement module and the modal enhancement strategy will get the best performance of 0.91635, which also shows that the two innovation points can be effective at the same time and contribute to the automatic personality recognition task.
\begin{center}
Table 4 Results of ablation experiments

\begin{tabular}{lllllll}

  \toprule
 Method &E&N&A&C&O&Aver\\
  \midrule
  Baseline&0.91635&	0.91105	&0.90788&	0.90821&	0.91153&	0.91100 \\
  \midrule
Baseline+ MSFEM&0.91784	&0.91201&	0.91121&	0.91170&	0.91261	&0.91308 \\
   \midrule
Baseline+MAS & 0.91991&	0.91576&	0.91425&	0.91576&	0.91533	&0.91620          \\
     \midrule
Baseline+ MSFEM+MAS  & 0.92005	&0.91594	&0.91335	&0.91753&	0.91489	&0.91635      \\

  \bottomrule
\end{tabular}
\end{center}
\section{Conclusion}
In this paper, we propose an end-to-end personality recognition framework for automatic perception of personality from audio and video. To address the difficulty of multimodal data, we design a multiscale feature augmentation module that applies to both visual and auditory modalities at the same time, and in order to solve the problem of model robustness, we introduce a modal enhancement strategy during the training process to simulate a variety of modal anomalies. The experimental results show that our proposed both enhancement module and the modal enhancement strategy have their own contributionsmultiscale feature , and the combined model achieves SOTA for personality recognition in audiovisual modality, and in this paper, we have simulated six non-ideal data scenarios as a test, and the results show that the model using the modal enhancement strategy is more robust. This paper focuses on audio-visual modality, however, textual modality is also more mainstream nowadays, especially in the era of big models, natural language processing technology has been qualitatively improved, so the follow-up work will consider combining textual modality to construct a tri-modal model. And in addition to the personality dataset itself, it is hoped to introduce other auxiliary information, such as emotional expressions to enhance the perception of personality.

\bibliographystyle{unsrt}  
\bibliography{references}

\end{document}